\documentclass[%
 aip,
 amsmath,amssymb,
 reprint,%
]{revtex4-1}

\usepackage{graphicx}
\usepackage{dcolumn}
\usepackage{bm}

\usepackage[utf8]{inputenc}
\usepackage[T1]{fontenc}
\usepackage{mathptmx}
\usepackage{etoolbox}
\usepackage{hyperref}
\hypersetup{
    colorlinks=true,
    linkcolor=black,
    filecolor=black,      
    urlcolor=black,
    citecolor=black,
    }
\usepackage{subcaption}
\setcitestyle{authoryear}
\usepackage[colorinlistoftodos]{todonotes}
\newcommand{\todURi}[1]%
{\todo[color=blue!10,inline]{{\bf Uli:} #1}}
\newcommand{\todUR}[1]%
{\todo[color=blue!10,linecolor=darkgray]{\footnotesize {\bf Uli:} #1}}

\usepackage{xspace}

\newcommand{\lbmpy}{{\em lbmpy}\xspace}

\newcommand{\walberla}{\textsc{waLBerla}\xspace}
\newcommand{\nvidia}{NVIDIA\xspace}

\makeatletter
\def\@email#1#2{%
 \endgroup
 \patchcmd{\titleblock@produce}
  {\frontmatter@RRAPformat}
  {\frontmatter@RRAPformat{\produce@RRAP{*#1\href{mailto:#2}{#2}}}\frontmatter@RRAPformat}
  {}{}
}%
\makeatother
\begin{document}

\preprint{AIP/123-QED}

\title[Large-Scale Simulations of Fully Resolved Complex Moving Geometries with Partially Saturated Cells]{Large-Scale Simulations of Fully Resolved Complex Moving Geometries with Partially Saturated Cells}
\author{P. Suffa}
\email{philipp.suffa@fau.de}
\affiliation{ 
Chair for System Simulation, Friedrich Alexander Universität Erlangen-Nürnberg, Erlangen, Germany
}
\author{S. Kemmler}
\affiliation{ 
Division 7.2, Bundesanstalt für Materialforschung und -prüfung (BAM), Berlin, Germany
}
\affiliation{ 
Chair for System Simulation, Friedrich Alexander Universität Erlangen-Nürnberg, Erlangen, Germany
}

\author{H. Koestler}
\affiliation{ 
Chair for System Simulation, Friedrich Alexander Universität Erlangen-Nürnberg, Erlangen, Germany
}
\author{U. Ruede}
\affiliation{ 
Chair for System Simulation, Friedrich Alexander Universität Erlangen-Nürnberg, Erlangen, Germany
}

\date{\today}

\begin{abstract}
We employ the Partially Saturated Cells Method (PSM) to model the interaction between the fluid flow and solid moving objects as an extension to the conventional lattice Boltzmann method. 
We introduce an efficient and accurate method for mapping complex moving geometries onto uniform Cartesian grids suitable for massively parallel processing.
A validation of the physical accuracy of the solid-fluid coupling and the proposed mapping of complex geometries is presented.
The implementation is integrated into the code generation pipeline of the waLBerla framework so that highly optimized kernels for CPU and GPU architectures become available.
We study the node-level performance of the automatically generated solver routines.
71\% of the peak performance can be achieved on CPU nodes and 86\% on GPU accelerated nodes.
Only a moderate overhead is observed for the processing of the solid-fluid coupling when compared to the fluids simulations without moving objects.  
Finally, a counter-rotating rotor is presented as a prototype industrial scenario, resulting in a mesh size
involving up to 4.3 billion fluid grid cells. 
For this scenario, excellent parallel efficiency is reported in a strong scaling study on up to 32,768 CPU cores on the LUMI-C supercomputer and on up to 1,024 NVIDIA A100 GPUs on the JUWELS Booster system. 
\end{abstract}

\maketitle
\section{Introduction}

The efficient simulation of complex fluid flow remains a subject of industrial interest and is thus
the topic of continued intensive research.
The solution of physical problems often necessitates substantial computational resources, which may be met by the collective processing power of numerous Central Processing Unit (CPU) cores or the advanced parallel processing capabilities of Graphical Processing Units (GPUs). 
The leading High-Performance Computing (HPC) systems in the Top500 list (\url{www.top500.org}) are operating in an exascale regime, capable of executing more than $10^{18}$ calculations in a single second.

The Lattice Boltzmann Method (LBM) is a well-established technique for accurately modeling the physical properties of fluid flow.
The LBM is essentially an explicit time-stepping scheme and thus has the advantage of being inherently parallel and well-suited for exploiting node-level and instruction-level parallelism. 
This makes the LBM a good candidate for processing on GPUs and for large-scale simulations on supercomputers (\cite{watanabePerformanceEvaluationLattice2022}, \cite{liuAcceleratingLargeScaleCFD2023}, \cite{spinelliHPCPerformanceStudy2023}, \cite{holzerHighlyEfficientLattice2021}). 

\begin{figure}
	\centering
     \includegraphics[width=\linewidth]{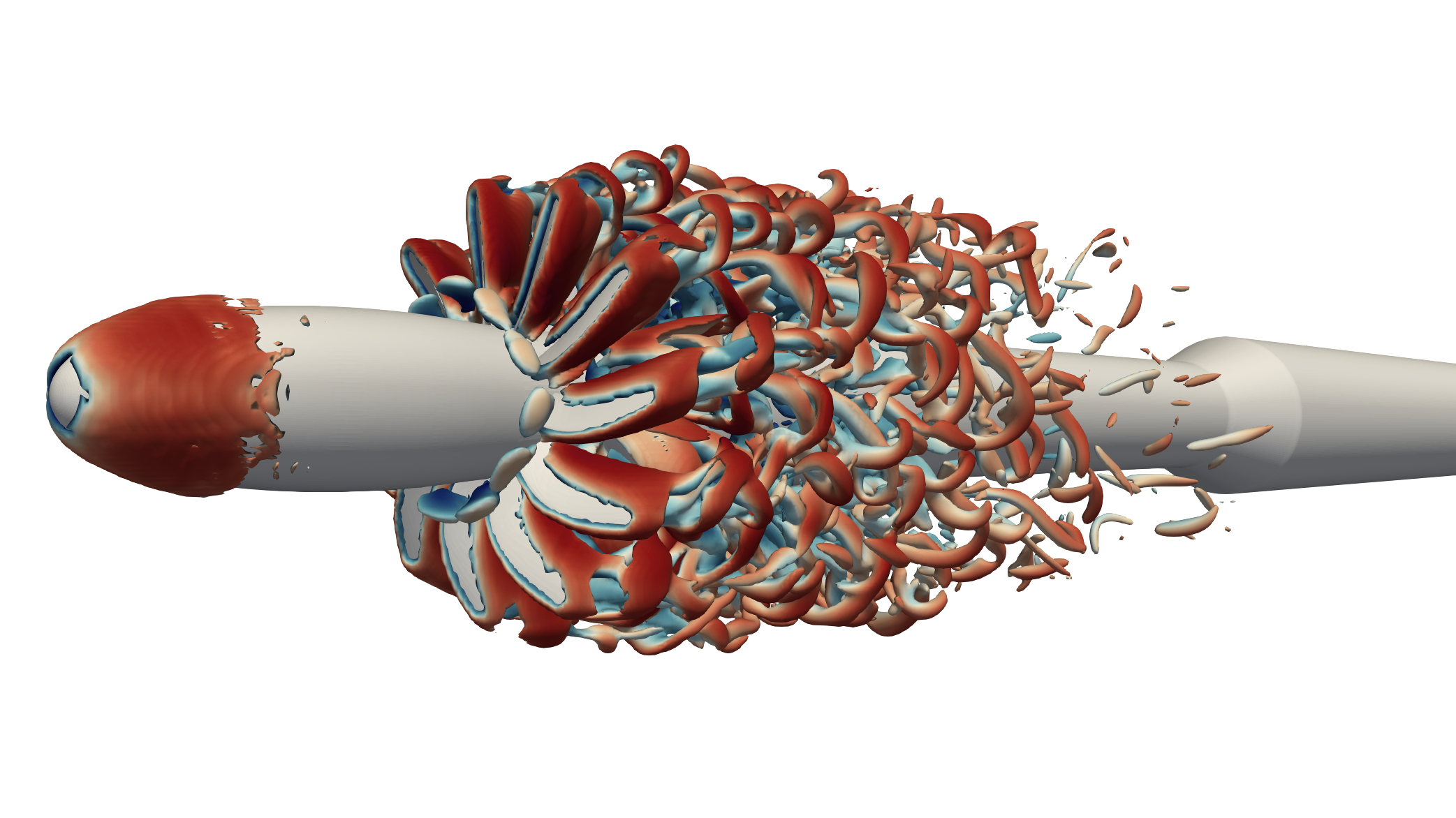}
    \caption{Visualization of the flow around a counter rotating open rotor by the Q-criterion of the flow. (Video available online)}
    \label{fig:CROR_big_run}
\end{figure}

Domains with moving boundaries play a significant role in computational fluid dynamics for many industrial applications.
These include complex geometries like wind turbines, aircraft propellers, and chemical mixers. 
Complex geometries often cannot be described analytically or as composed of simple basic shapes.
This poses a significant challenge in efficiently mapping such geometries onto a uniform Cartesian grid as it is used for the LBM, particularly when the geometry field representation must be updated every time step to accommodate a moving geometry.
The present study uses a counter-rotating open rotor (CROR), as illustrated in \autoref{fig:CROR_big_run}, as a prototype scenario to study complex and typical flow phenomena and algorithms and methods needed for their accurate modeling.
The renewed interest in open rotors as prospective propulsion units stems from their ability to reduce emissions of carbon dioxide ($CO_2$) and nitrogen oxides ($NO_x$) (\cite{lyuSlidingMeshApproach2022}). 
The simulation of flows around complex moving geometries, such as the CROR, necessitates an efficient and accurate coupling method to facilitate momentum exchange between solid and fluid domains. 
Various approaches have been developed to address this challenge in LBM simulations.\newline
One such variant is the Momentum Exchange Method (MEM), which was originally proposed by \cite{laddNumericalSimulationsParticulate1994}. 
The MEM addresses the coupling from solid to fluid by boundary conditions, such as a simple No-Slip boundary, but also higher-order boundary conditions are possible (\cite{Ginzburg2008427}).
The fluid-to-solid coupling is achieved by calculating the acting force and torque on the solid object by summing up all probability distribution functions (PDFs) pointing toward the surface of the object.
Due to an explicit mapping of the object into the domain, grid cells will transition from solid to fluid when objects move. 
Thus, suitable fluid information must be reconstructed on the affected cells by an appropriate 
methods to compute the PDFs. 
This is sometimes called the {\em fresh-node} problem.
The MEM has been demonstrated to exhibit good accuracy in the context of fluid-solid interactions (\cite{rettingerEfficientFourwayCoupled2022}). 
However, due to the relatively complex inherent logic, it appears to be difficult to implement efficiently on current GPUs.

An alternative approach to model the fluid-structure interaction is the Immersed Boundary Method (IBM). 
This method was first introduced to LBM by \cite{fengImmersedBoundarylatticeBoltzmann2004}.
The IBM approximates a boundary by a set of off-lattice marker points that affect the fluid only via a force field. 
The advantages of this approach include the fact that the solid object does not need to be mapped explicitly onto the LBM lattice. 
Additionally, this approach facilitates the realization of moving and deformable boundaries without the necessity of re-meshing.
A notable disadvantage of IBM is the intricate challenge of distributing the force points efficiently onto a three-dimensional geometry (\cite{lbm_book}). 
Moreover, IBM is reported to be only a first-order accurate boundary condition for sharp interfaces (\cite{peskinImmersedBoundaryMethod2002}). 

The sliding mesh approach is another viable method for simulating rotating objects in a simulation domain. 
This approach involves the division of the computational domain into an inner and an outer lattice. 
During the simulation of a rotating object, the inner lattice rotates in unison with the object, while the outer frame remains stationary. 
The transition between these grids is facilitated by an overlapping interface and an interpolation procedure between the two grids (\cite{zhang2011}). 
The authors of \cite{lyuSlidingMeshApproach2022} have successfully presented a sliding mesh approach for the simulation of an open rotor. 
A key advantage of this approach is that it requires only a single mapping of the geometry onto the lattice at the initiation of the simulation. 
However, this approach is not without its limitations, including constraints on rotating geometries, and the potential for challenges in efficient implementations on GPU architectures.

It should be noted that further alternative approaches, such as the actuator line and actuator disc models, have been developed. 
However, these models do not resolve the underlying complex geometry and their boundary conditions; rather, they are based on modeling the forces acting on the fluid. 
These methodologies are cost-effective and are frequently employed for wind turbine simulations, as evidenced in \cite{schottenhammlWaLBerlawindLatticeBoltzmannbasedHighperformance2024} and \cite{rullaudActuatorLineModelLattice2018}.
In \cite{ribeiroSlidingMeshSimulations2024}, the actuator line model is combined with a sliding mesh approach to also simulate wind turbines.

In this study, we employ and study the Partially Saturated Cells Method (PSM) for representing fully-resolved complex moving geometries in LBM solvers. 
The PSM was first introduced by \cite{nobleLatticeBoltzmannMethodPartially1998a} and is commonly used for fluid-particle simulations with geometrically resolved particles (\cite{majumderReexaminingPartiallySaturatedcells2022a}, \cite{rettingerComparativeStudyFluidparticle2017a}). 
The PSM is naturally parallel since it is structurally very similar to the pure LBM.
Thus it is in particular a suitable choice for simulations on GPUs that require an explicit fluid-solid interaction (\cite{kemmlerEfficiencyScalabilityFullyresolved2025}, \cite{benseghierParallelGPUbasedComputational2020}). 
The authors of \cite{trunkSimulationArbitrarilyShaped2018} demonstrated, that the PSM is not constrained to spherical particles, but can handle arbitrarily shaped objects.
A notable advantage of the PSM is that it does not require explicit boundary handling to model solid objects in the domain. 
The no-slip boundary condition on the object surface is implicitly integrated into the LBM step itself. 
Furthermore, when using the PSM, there is no need for reconstructing valid fluid information in cells that transition from a solid to a fluid state.

The remaining article is structured as follows.
In \autoref{sec:psm_theory} we present the theory of the PSM and the integration into the code generation pipeline of \lbmpy{} (\cite{lbmpy}).
In \autoref{sec:mapping} we present an efficient way of mapping a complex geometry into an LBM lattice in every time step to handle the rotation/translation of the solid object.
The accuracy of the generated PSM is validated using the standard test of a settling sphere in \autoref{sec:validation}.
In \autoref{sec:performance}, the industrial test case of the CROR is set up with the large-scale multi-physics framework \walberla{} (\cite{waLBerla}). 
The node-level performance and the scaling efficiency of the simulation are evaluated.
The performance tests are conducted on two distinct platforms: LUMI-C (\url{www.lumi-supercomputer.eu}), a large CPU-based supercomputing system currently listed at rank 8 on the TOP500 list; and JUWELS Booster (\cite{juwels}), a large-scale GPU accelerated system.

\section{The Partially Saturated Cells Method}
\label{sec:psm_theory}

\subsection{Theory}
The PSM can be regarded as an extension of the LBM, which incorporates momentum exchange between the solid and fluid phases of the simulation.
The LBM is based on Cartesian grids, and each grid cell contains PDFs that are updated in a time-stepping scheme based on the Boltzmann equations.
The lattice Boltzmann update step is expressed as
\begin{equation}
    f_i(\textbf{x} + \Delta t \textbf{c}_i, t+ \Delta t) = f_i(\textbf{x},t) + \Omega_i^F, 
\end{equation}
where $f_i(\textbf{x},t)$ is the PDF in cell $\textbf{x}$ corresponding to the stencil direction $i$, and $\textbf{c}_i$ is the discrete lattice velocity in $i$. The collision operator $\Omega_i^F$ relaxes the PDFs towards their equilibrium in the Boltzmann equations.
The most basic collision operator is the Single Relaxation Time operator (SRT) and reads 
\begin{equation}
    \Omega_i^F = - \frac{1}{\tau}(f_i(\textbf{x},t) - f^{(\text{eq})}_i(\textbf{x},t)), 
\end{equation}
with $\tau$ as the relaxation time for the LBM. 
We present in \autoref{sec:code_gen} that more complex collision operators can also be combined with the PSM.
The equilibrium function yields 
\begin{equation}
    f_i^{(\text{eq})}(\textbf{u}, \rho) = \omega_i \rho \left[1 + \left( \frac{\textbf{c}_i \cdot \textbf{u}}{c^2_s} + \frac{(\textbf{c}_i \cdot \textbf{u})^2}{2 c^4_s} + \frac{\textbf{u}^2}{2 c^2_s} \right) \right],
\end{equation}
with weighting factor $\omega$ for stencil direction $i$, fluid velocity $\textbf{u}$, fluid density $\rho$, and lattice speed of sound $c_s = \sqrt{1/3}$.

The lattice Boltzmann equation can be extended by a solid collision operator $\Omega_i^S$ and a solid volume fraction field $B(\textbf{x}, t)$ to result in the equation for the PSM as
\begin{multline}
        f_i(\textbf{x} + \Delta t \textbf{c}_i, t + \Delta t) = \\ f_i(\textbf{x},t) + (1 - B(\textbf{x}, t)) \cdot \Omega_i^F + B(\textbf{x}, t) \cdot \Omega_i^S.
        \label{equ:PSM_rule}
\end{multline}
Here, $B(\textbf{x}, t)$ is a scalar field indicating the presence or absence of a solid object in a specific cell. 

Two approaches exist for calculating $B(\textbf{x}, t)$ from $\epsilon(\textbf{x}, t)$. 
$\epsilon(\textbf{x}, t)$ is the overlap of the solid object and an LBM lattice cell $\textbf{x}$, with $\epsilon(\textbf{x}, t) = 1$ for a cell fully covered by the solid object, and $\epsilon(\textbf{x}, t) = 0$ for a cell not interacting with the object. 
The straightforward approach is a direct mapping, so
\begin{equation}
    B(\textbf{x}, t) = \epsilon(\textbf{x}, t).
\end{equation}

The authors of \cite{rettingerComparativeStudyFluidparticle2017a} report better accuracy when weighting  $\epsilon(\textbf{x}, t)$ with the relaxation time $\tau$ to yield 

\begin{equation}
    B(\textbf{x}, t) = \frac{\epsilon(\textbf{x}, t) \left(\tau - \frac{1}{2}\right)}{\left(1 - \epsilon(\textbf{x}, t)\right) + \left(\tau - \frac{1}{2}\right)}.
\end{equation}

\begin{figure}
\centering
\includegraphics[width=0.85\linewidth, trim={10cm 9.5cm 13cm 2.5cm},clip]{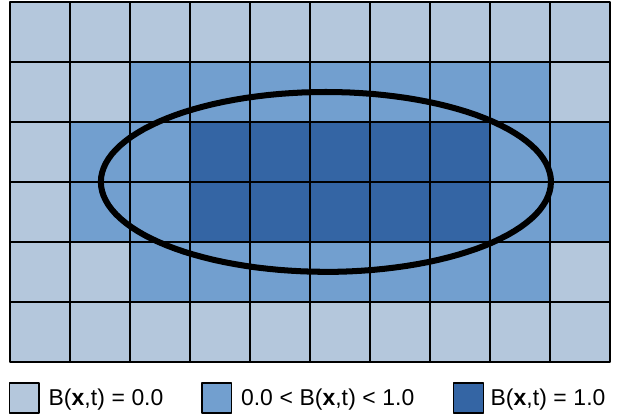}
\caption{\label{fig:fractionField} The fraction field for the Partially Saturated Cells Method representing an example geometry of an ellipse.}
\end{figure}

In \autoref{fig:fractionField}, the volume fraction field of an ellipse is displayed. 
Cells lying outside of the ellipse shape have a solid volume fraction $B(\textbf{x}, t) = 0$, cells completely inside the ellipse boundaries correspond to a fraction value of $B(\textbf{x}, t) = 1$, 
and cells at the interface have a value between 0 and 1, depending on the overlapping volume of the cell with the object. 
Computing this fraction field efficiently becomes challenging in the PSM (\cite{lbm_book}), 
especially if the objects are moving, and thus, the volume fraction field must be updated frequently.
In \autoref{sec:mapping}, we will present in detail how the volume fraction field is initialized and updated in an efficient way for a moving geometry.

The solid collision operator $\Omega_i^S$ in \autoref{equ:PSM_rule} functions as a No-Slip boundary condition.
According to \cite{haussmannGalileanInvarianceStudy2020}, it can have the three different appearances
\begin{multline}
    \text{SC1: } \Omega_i^S = \\ \left[f_{\bar{i}}(\textbf{x},t) - f_{\bar{i}}^{(\text{eq})}(\rho, \textbf{u}) \right] - \left[f_i(\textbf{x},t) - f_i^{(\text{eq})}(\rho, \textbf{u}_s) \right],
    \label{equ:solid_collision1}
\end{multline}
\begin{multline}
    \text{SC2: } \Omega_i^S = \\ \left[f_i^{(\text{eq})}(\rho, \textbf{u}_s) - f_i(\textbf{x},t) \right] + \left( 1 - \frac{1}{\tau} \right) \left[ f_i(\textbf{x},t) - f_i^{(\text{eq})}(\rho,\textbf{u}_s \right],
    \label{equ:solid_collision2}
\end{multline}
\begin{multline}
    \text{SC3: } \Omega_i^S = \\ \left[f_{\bar{i}}(\textbf{x},t) - f_{\bar{i}}^{(\text{eq})}(\rho, \textbf{u}_s) \right] - \left[f_i(\textbf{x},t) - f_i^{(\text{eq})}(\rho, \textbf{u}_s) \right].
    \label{equ:solid_collision3}
\end{multline}

In these equations, $\bar{i}$ is the opposite PDF direction of $i$, and $\textbf{u}_s$ refers to the velocity of the solid volume. 
According to \autoref{equ:PSM_rule}, a solid volume fraction $ B(\textbf{x}, t) = 0$ results in a pure LBM collision, while  $ B(\textbf{x}, t) = 1$ describes a bounce-back of the equilibrium. 
For partially saturated cells with $ 0 < B(\textbf{x}, t) < 1$, a mixed collision and bounce-back scheme is performed.
This approach enables the capture of the influence of a solid volume on the flow field. 
The coupling in the other direction, from the flow field to the solid volume, is given by the hydrodynamic forcing term
\begin{equation}
    \textbf{F}(t) = \frac{\Delta x^3}{\Delta t} \sum_{\textbf{x}_s} B(\textbf{x}_s, t) \sum_i \Omega_i^S \textbf{c}_i ,
    \label{equ:forcing}
\end{equation}
and the torque term
\begin{equation}
    \textbf{T}(t) = \frac{\Delta x^3}{\Delta t} \sum_{\textbf{x}_s}  B(\textbf{x}_s, t) (\textbf{x}_s - \textbf{R}) \times \sum_i \Omega_i^S \textbf{c}_{i}, 
    \label{equ:torque}
\end{equation}
with $\textbf{x}_s$ as all cells in contact with the object and $\textbf{R}$ as the center of mass of the solid object (\cite{lbm_book}).

The No-Slip condition is directly incorporated into the PSM update rule in \autoref{equ:PSM_rule}.
Thus, no special boundary handling is needed for No-Slip boundaries. 
This improves computational performance.
Additionally, the PSM is reported to be perfectly mass conserving (\cite{lbm_book}) and second-order accurate (\cite{rettingerComparativeStudyFluidparticle2017a}, \cite{nobleLatticeBoltzmannMethodPartially1998a}).
A notable benefit of the PSM is the elimination of the fresh-node problem. 
In the context of LBM simulations involving moving solid objects, a scenario inevitably arises where the moving object ceases to cover a specific cell. 
In the context of direct mapping for fluid-solid coupling, this cell previously possessed a boundary state, so it did not contain valid PDF information.
When transforming to a fluid state, the cell must be re-initialized with a valid set of PDFs, and therefore, some PDF reconstruction algorithm is required (fresh-node problem).
This needs to be handled carefully (\cite{rettingerEfficientFourwayCoupled2022}).
In the PSM, cells (partially) covered by a solid object also participate in the collision (\autoref{equ:PSM_rule}) and, therefore, always hold valid PDF information, so no explicit PDF reconstruction is needed. 
This property, combined with the PSM's inherent parallelism, makes it well-suited for executing moving body simulations on GPU accelerators.

\subsection{Code Generation for Efficient LB Methods}
\label{sec:code_gen}
\begin{figure}
	\centering
    \includegraphics[width=0.7\linewidth]{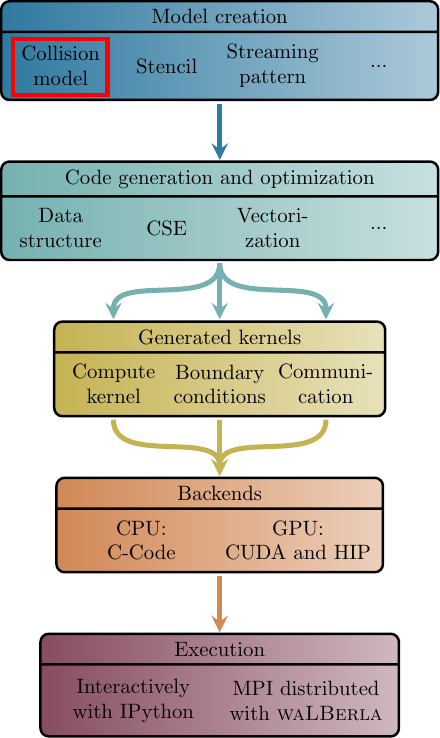}
	\caption{Complete workflow of the code generation pipeline of \lbmpy{}. The code generation for the PSM is integrated in the collision model of the model creation stage. (Figure from \cite{suffaArchitectureSpecificGeneration2024}).}
	\label{fig:lbmpywithwalberla}
\end{figure}

Here, we will describe our in-house code generation framework \lbmpy{} (\cite{lbmpy}) and how it is employed and extended.
The automatic code generation can produce particularly efficient implementations of the LBM,
suitable for large-scale parallel fluid simulations.
This is achieved with our multi-physics framework \walberla{} (\cite{waLBerla}) in which the automatically generated code fragments are integrated.
An overview of the code generation process is presented in \autoref{fig:lbmpywithwalberla}.

In the model creation stage, a user of \lbmpy{} must concisely describe the LBM's configuration. 
A variety of LBM variants are available. 
In particular, a range of collision operators with increasing complexity exist, including SRT, TRT, MRT, Central Moments, or the Cumulant operator (\cite{lbm_book}, \cite{GEIER2015507}).
Furthermore, LBMs can be configured with diverse streaming patterns, such as the two-field pattern with either a push or pull (\cite{lbm_book}) semantics.
To reduce memory consumption and the number of memory accesses, in-place streaming patterns such as the AA-pattern (\cite{baileyAcceleratingLatticeBoltzmann2009}) and the Esotwist streaming pattern (\cite{geierEsotericTwistEfficient2017}) can be used.
\lbmpy{} supports all these collision operators and streaming patterns.
It further supports different velocity stencils, including D2Q9, D3Q15, D3Q19, and D3Q27, as well as various forcing and turbulence models.

After the user's model definition, the actual code generation is executed based on specific optimization parameters. 
This includes the selection of a data structure that can be directly or indirectly addressed (\cite{suffaArchitectureSpecificGeneration2024}). 
Other optimizations include common sub-expression elimination, which reduces the number of FLOPS in the generated code, and vectorization, which supports vector instruction sets such as SSE, AVX, and AVX512 on modern CPUs. 

\lbmpy{} generates efficient code for the LBM compute kernels and 
-- equally importantly -- also for the boundary conditions and the parallel communication routines.  
By supporting different backends, \lbmpy{} can generate code for most CPU types and AMD and NVIDIA GPUs. 
For CPUs, the generated output is plain C code;
for NVIDIA GPUs, the output is CUDA; for AMD GPUs, it is HIP code.
For a comprehensive overview of the code generation pipeline's functionalities, refer to \cite{lbmpy}, \cite{Bauer19}, \cite{hennigAdvancedAutomaticCode2022}, and the open-source code repository (\url{https://i10git.cs.fau.de/pycodegen/lbmpy}).

The PSM equations have been integrated into the code generation of \lbmpy{}. 
In the collision model creation, the fluid collision operator is weighted by $(1 - B(\textbf{x}, t))$ and the solid collision multiplied by the solid volume fraction is added to the term to result in \autoref{equ:PSM_rule}. 
All three solid collision operators from \autoref{equ:solid_collision1}, \autoref{equ:solid_collision2} and \autoref{equ:solid_collision3} were integrated.
Finally, the force and torque equation in \autoref{equ:forcing} and \autoref{equ:torque} is implemented.
With these modifications, \lbmpy{} can now generate PSM kernels combined with the complete choice of all collision operators and streaming patterns. 
Of course, all optimizations of \lbmpy{} can be invoked and lead to excellent node-level efficiency, as presented in the following. 
Most importantly, this new extension of \lbmpy{} enables the execution of the PSM on multiple hardware platforms, including all common CPUs and NVIDIA and AMD GPUs.

\section{Efficient Mapping of Complex Moving Geometries}
\label{sec:mapping}

\begin{figure*}
	\centering
     \includegraphics[width=\linewidth, trim={0cm 1cm 0cm 2cm},clip]{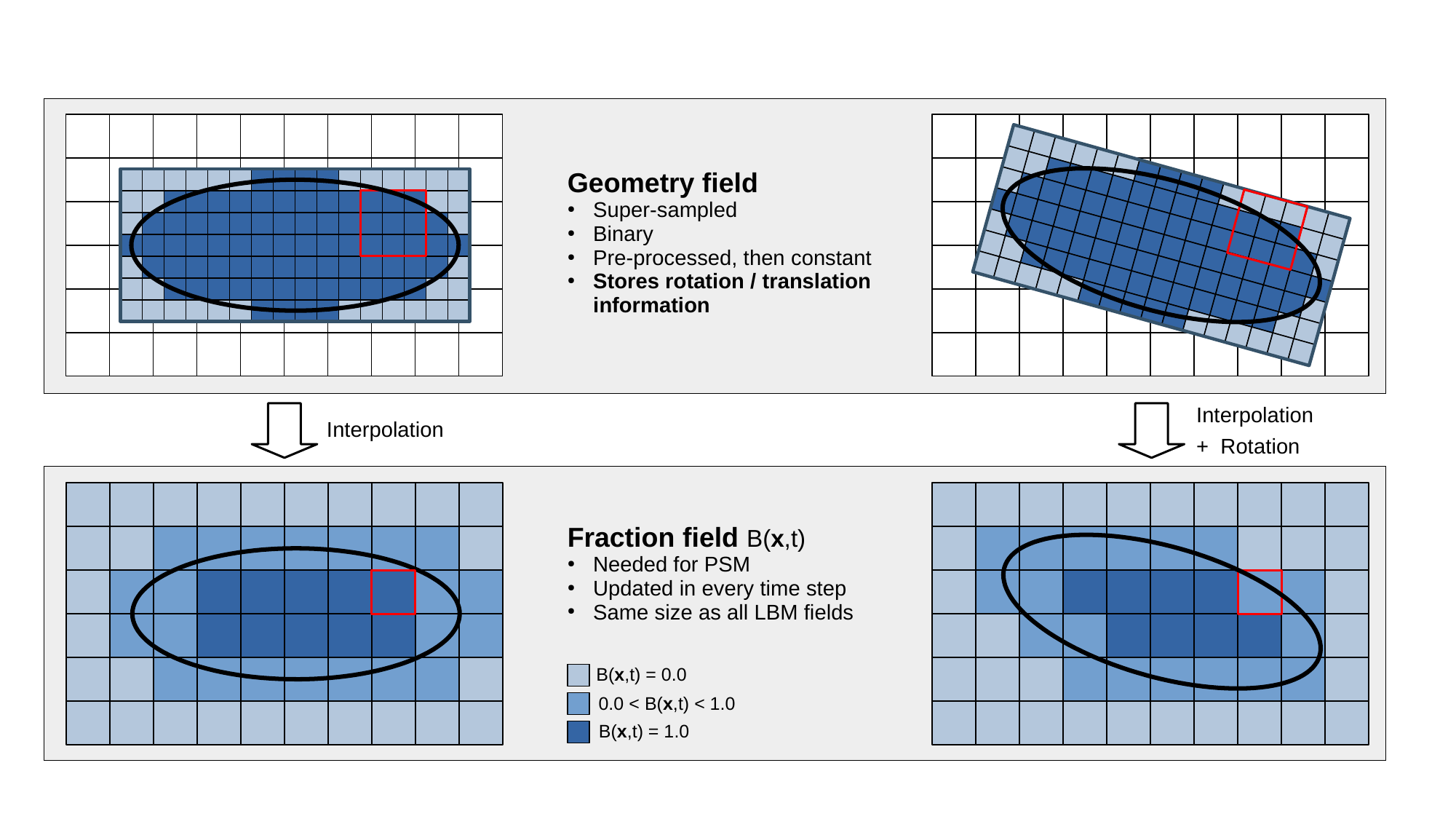}
    \caption{Interplay of the fraction field and the super-sampled geometry field for an oval shape with super-sampling depth of 1.}
    \label{fig:geometryField}
\end{figure*}

In the process of setting up a complex domain for a fluid dynamics simulation, a description of the geometry of interest is given. 
In instances where the geometry can be described analytically -- for instance, a sphere or a cylinder -- the mapping of the object onto the LBM lattice is both straightforward and cheap.
Conversely, the mapping process is complicated when the geometry is complex and thus cannot be described by basic shapes.

\begin{figure}
	\centering
     \includegraphics[width=\linewidth, trim={0cm 9.5cm 21.5cm 0cm},clip]{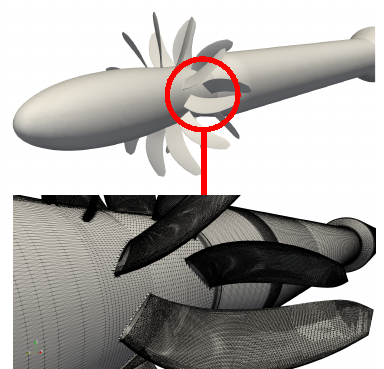}
    \caption{Counter rotating open rotor geometry with 610,692 vertices and 1,221,380 faces. The front rotor rotating clockwise, while the back rotor (stator) rotating counter-clockwise.}
    \label{fig:CROR_geo}
\end{figure}

In \autoref{fig:CROR_geo}, the geometry of a counter-rotating open rotor is presented. 
The figure reveals the complex structure of its vertices and faces. 
The geometry information consists of over 600.000 vertices and over 1.200.000 faces. 
As discussed in \autoref{sec:psm_theory}, a crucial part of the PSM algorithm is the construction of the solid volume fraction field $B(\textbf{x}, t)$.
To achieve this objective, we have to map the complex geometry, which consists of vertex and face information, to the uniform lattice of the fraction field. 
This process is referred to as voxelization. 
Voxelization can be costly, depending on the complexity of the geometry. 
This process can be manageable as a pre-processing step if the geometry remains stationary. 
However, it may be necessary to voxelize the geometry continuously when dealing with moving geometry. 
Given the substantial number of vertices and faces, efficiently rotate the vertices and map/voxelize the geometry onto the fraction field repeatedly is not a viable option. 

Therefore, we introduce a method to handle moving geometries that only needs to voxelize the geometry once. 
This approach draws inspiration from the methodology outlined in \cite{trunkSimulationArbitrarilyShaped2018}.
The methodology involves the creation of a second field, designated as the \textit{geometry field}.
This field is a super-sampled representation of the fraction field. 
It is only stored in the area around the geometry, and its data type is binary, indicating whether a super-sampled cell center is within or outside the geometry. 
This information is introduced by voxelizing the geometry onto the geometry field once as a pre-processing step.
An illustration of an artificial geometry field is provided in \autoref{fig:geometryField}, which depicts the geometry field of an ellipse with a mesh spacing of $\Delta x_{\text{geo}} = \Delta x_{\text{LBM}}/2$, so the resolution of the geometry field $\Delta x_{\text{geo}}$ is double the resolution of the solid volume fraction field (LBM lattice) $\Delta x_{\text{LBM}}$ in every dimension. 
The super-sampling parameter $s$ is a parameter that enhances the precision of the geometry representation by resolving the geometry field with an adjustable number of sub-cells, although this results in an increase in memory consumption. 
A factor of $s$ results in $2^s$ super-sampled cells per direction and ${2^s}^3$ cells per LBM cell.
Therefore, setting $s=1$ leads to 8 cells per LBM cell for a 3D simulation, while $s=2$ results in 64 super-sampled cells.

With the geometry field in hand, which yields an accurate representation of the actual geometry, we can now create the desired fraction field. 
This is achieved by interpolating from the geometry field to the fraction field.
For illustrative purposes, we direct our attention to one of the red-marked cells located in the bottom of \autoref{fig:geometryField}. 
To get the information of the cell's solid volume fraction, we average over all corresponding super-sampled cells of the geometry field (upper figures).

We do not have to explicitly rotate vertices or cells when translating or rotating the geometry.
Instead, the rotation and/or translation information is incorporated while identifying the corresponding cells in the geometry field, thereby acquiring the rotated information on the solid volume fraction field.
This is achieved by multiplying the cell center in LBM space with a rotation matrix and/or adding a translation vector to get the translated/rotated cell in geometry space. 
Consequently, the geometry's movement is implicitly managed without introducing additional overhead.

In the context of GPU execution, the geometry field is initially populated on the CPU and subsequently transferred to the GPU once. 
The interpolation kernel, which constructs the solid volume fraction field from the geometry field at each temporal step, is executed entirely on the GPU.
In \autoref{sec:nodePerformance}, we will demonstrate that the interpolation process is remarkably efficient.
So, utilizing this method enables the efficient management of movement for fully resolved complex geometries on an LBM lattice.

\subsection{Verification of the Geometry Mapping}

\begin{figure*}
	\centering
      \begin{subfigure}{0.37\textwidth}
         \centering
         \includegraphics[width=0.7\linewidth,trim={12cm 3cm 12cm 3cm},clip]{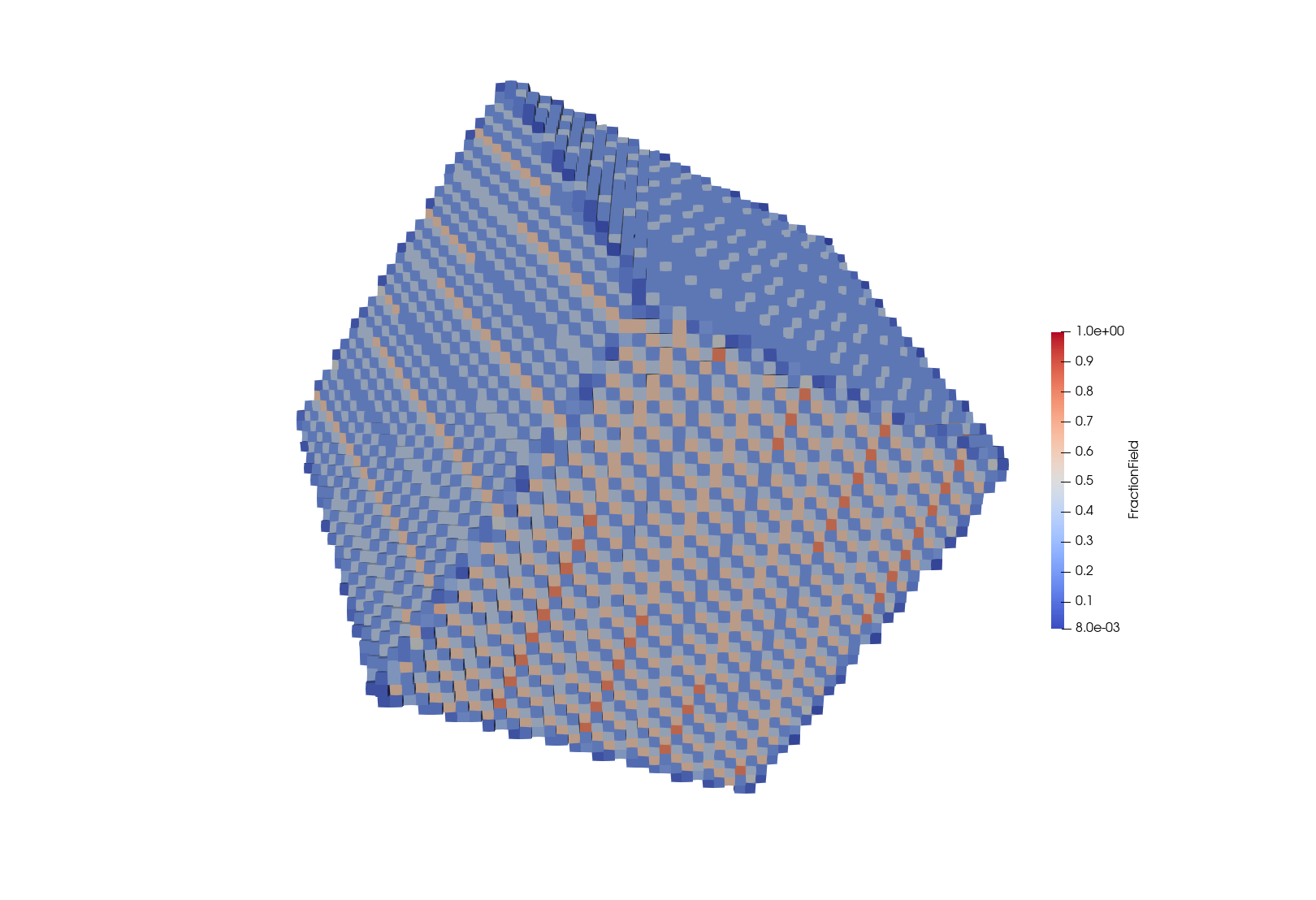}
         \subcaption{Fraction field of a rotating cube geometry with $N=20$ and $s=2$.}
         \label{fig:cube}
     \end{subfigure}
     \begin{subfigure}{0.3\textwidth}
         \centering
         \includegraphics[width=0.7\linewidth,trim={12cm 3cm 12cm 3cm},clip]{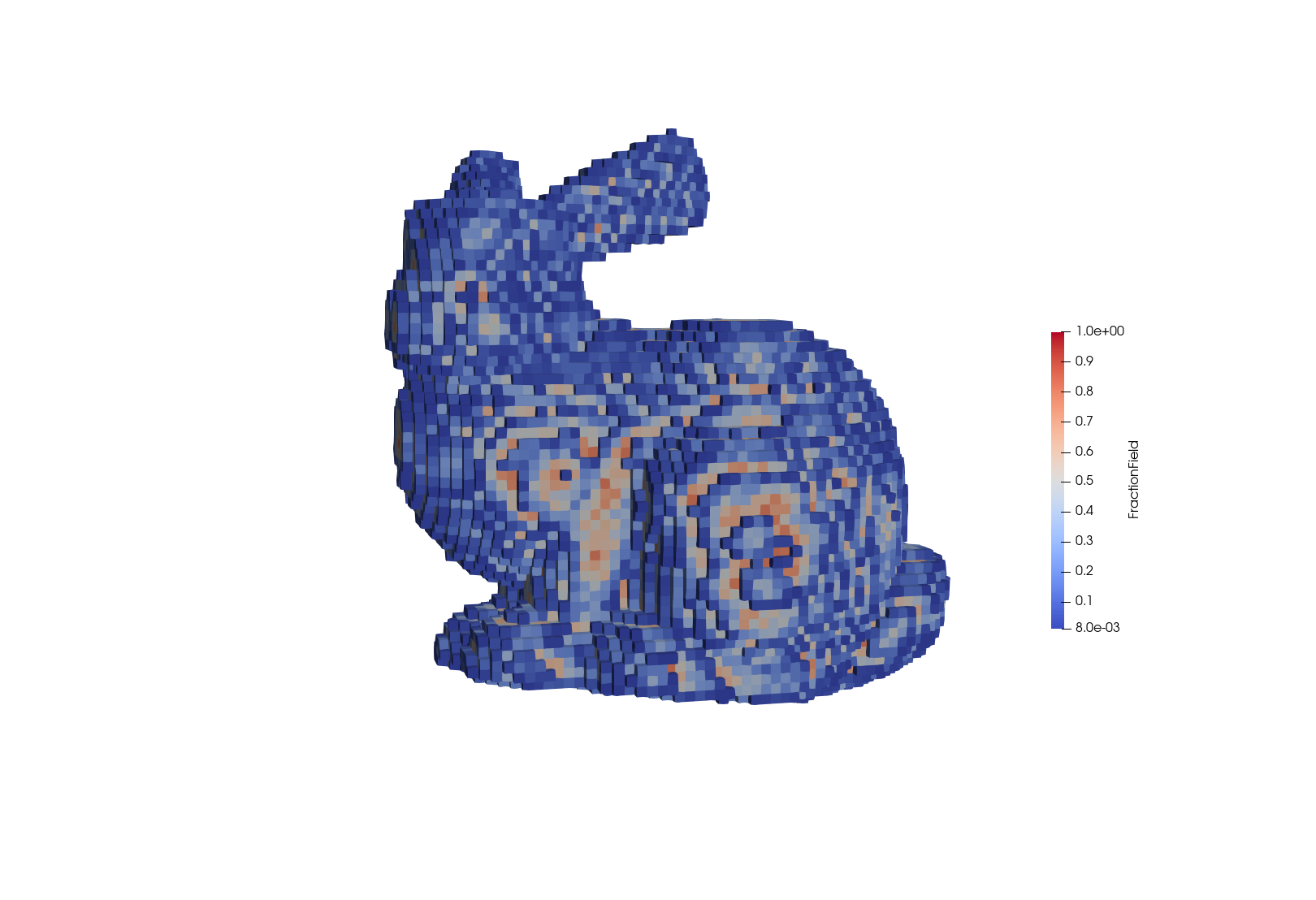}
         \subcaption{Fraction field of a rotating Stanford bunny with $N=20$ and $s=2$.}
         \label{fig:bunny}
     \end{subfigure}
     \begin{subfigure}{0.3\textwidth}
         \centering
         \includegraphics[width=0.7\linewidth,trim={12cm 3cm 12cm 3cm},clip]{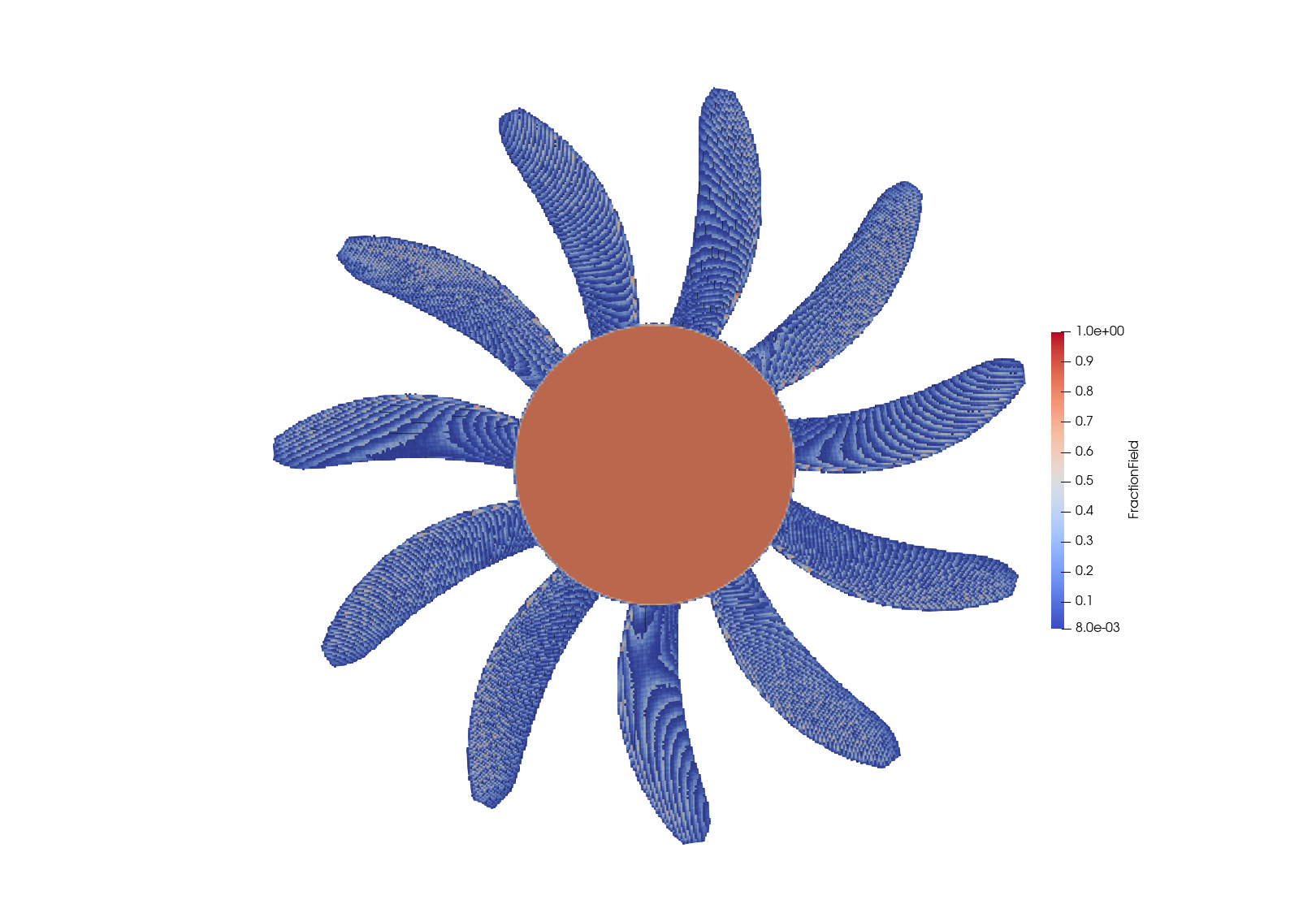}
         \subcaption{Fraction field of a rotating rotor with $N=20$ and $s=2$.}
         \label{fig:rotor}
     \end{subfigure}
    \caption{Visualization of the fraction field for three different geometries. A reddish color represents a fraction value close to 1 (cell fully covered by the object), a bluish color represents a fraction field close to 0 (cell not covered by the object). }
	\label{fig:meshes}
\end{figure*}

The validity of the geometry mapping is confirmed through a comparison of the solid fraction volume
\begin{equation}
    V_s(t) = \sum_{\textbf{x}_s} B(\textbf{x}, t) \Delta x,
    \label{equ:fractionVolume}
\end{equation}
averaged in time with the actual volume of an object using the L2-error norm.
The verification process involves the quantification of the volume of a rotating cube, a geometry mesh representing a bunny (Stanford bunny from \url{http://graphics.stanford.edu/data/3Dscanrep/}), and one rotor of our CROR geometry as illustrated in \autoref{fig:meshes}. 
We tested three resolutions for the LBM grid, namely $N=10$, $20$, and $40$, with $N$ denoting the number of cells in one dimension to resolve the geometry. 
Additionally, we investigated four super-sampling factors, namely $s=0$, $1$, $2$, and $3$.
We rotate the geometries for 100 time steps and calculate the average volume from \autoref{equ:fractionVolume}.

In the first verification case, a cube is rotating around all three axes to obtain various mappings of the cube onto the lattice.
\autoref{tab:cube_error} shows the L2-error of the resulting fraction field for the different parameters of the mesh resolution and the super-sampling factor. The error is averaged over 100 time steps.
For a factor $s = 0$ ($\Delta x_{\text{LBM}} = \Delta x_{\text{geo}}$), the volume error for the cube is comparable high for all resolutions. 
We also observe that for a resolution of $N=10$, the error magnitude can not be drastically reduced by increasing $s$.
So, a suitable configuration is a resolution of $N>10$ and $s>0$.
For a higher resolution of the LBM grid, so $N=40$, a lower super-sampling factor of $s=1$ or $s=2$ is sufficient.

In the second case using the rotating Stanford bunny geometry, the volume errors for all cases are approximately two magnitudes higher than for the cube, as presented in \autoref{tab:bunny_error}. 
This discrepancy may result from the more complex structure of the geometry, including convex and concave surfaces that are challenging to represent on a uniform lattice. 
Nevertheless, the behavior of the volume errors is similar to the cube geometry. 
By increasing $N$, the error decreases. 
The same holds for increasing $s$. 

The final verification case involves the rotor, which rotates around its z-axis. 
The resolution parameter $N$ quantifies the resolution in the z-direction. 
Consequently, the rotor is resolved at a factor of ten times higher in the x- and y-directions compared to the cube and the bunny, respectively, as the domain is ten times larger in these directions. 
The volume errors of the rotor are presented in \autoref{tab:rotor_error}. 
A super-sampling factor of 0 appears to be an inadequate selection for the rotor, resulting in substantial volume errors. 
This may be attributed to the thin nature of the rotor blades, which are challenging to capture due to the geometry field's low resolution. 
However, for higher LBM lattice resolution and higher $s$, a substantial decrease in the volume error is observed. 
For a resolution of $N=40$ and $s=3$, the memory consumption of the geometry field exceeded the memory of a modern CPU. Therefore, no data is presented. 

The implementation allows for a custom parametrization of $s$, enabling the user to trade off geometry representation precision for memory consumption.  
The grid resolution $N$ for the following simulation set-ups is naturally larger than $N=10$; therefore, the geometry representation on the uniform lattice of the solid volume fraction field for a super-sampling factor of $s>0$ is sufficient for all test cases.
A super-sampling factor of $s=1$ or $s=2$ is chosen for the following experiments. 
In \autoref{sec:nodePerformance}, we show that the factor $s$ has a low influence on the performance of the interpolation from the geometry field to the fraction field.

\begin{table}
\caption{L2 error of the averaged solid volume fraction for the cube geometry in \autoref{fig:cube}.}
\label{tab:cube_error}
\vspace{-0.3cm}
\begin{tabular}{ |c|c|c|c|c| } 
\hline
 N \textbackslash \ s & 0 & 1 & 2 & 3 \\ \hline
10	& 1.34E-05	& 9.06E-06	& 3.71E-06	& 1.65E-06 \\ \hline
20	& 5.35E-05	& 1.44E-07	& 5.03E-08	& 1.78E-08 \\ \hline
40	& 6.29E-06	& 8.94E-09	& 2.48E-09	& 1.03E-09 \\ \hline
\end{tabular}
\vspace{0.4cm}
\caption{L2 error of the averaged solid volume fraction for the Stanford bunny geometry in \autoref{fig:bunny}.}
\label{tab:bunny_error}
\vspace{-0.3cm}
\begin{tabular}{ |c|c|c|c|c| } 
\hline
 N \textbackslash \ s & 0 & 1 & 2 & 3 \\ \hline
10	& 4.17E-03	& 5.00E-05	& 3.59E-05	& 9.44E-06 \\ \hline
20	& 3.11E-05	& 1.86E-05	& 2.97E-06	& 7.48E-07 \\ \hline
40	& 1.63E-05	& 1.35E-06	& 4.35E-07	& 1.71E-07 \\ \hline
\end{tabular}
\vspace{0.4cm}
\caption{L2 error of the averaged solid volume fraction for the rotor geometry in \autoref{fig:rotor}.}
\label{tab:rotor_error}
\vspace{-0.3cm}
\begin{tabular}{ |c|c|c|c|c| } 
\hline
 N \textbackslash \  s & 0 & 1 & 2 & 3 \\ \hline
10	& 1.06E-05	& 7.79E-07	& 5.94E-07	& 1.50E-07 \\ \hline
20	& 6.62E-04	& 1.24E-07	& 3.29E-08	& 1.06E-08 \\ \hline
40	& 4.97E-04	& 6.06E-09	& 3.22E-09	& -        \\ \hline
\end{tabular}
\end{table}

\section{Validation of the Partially Saturated Cells Method for Moving Geometries}
\label{sec:validation}

The accuracy of the PSM has been studied in the following literature.
\cite{nobleLatticeBoltzmannMethodPartially1998a} showed second-order accuracy of the PSM for a moving cylinder setup. 
\cite{rettingerComparativeStudyFluidparticle2017a} validated the force acting on a periodic array of spheres and achieved second-order accuracy for the PSM. 
In \cite{majumderReexaminingPartiallySaturatedcells2022a}, the authors validate the PSM for the cases of a stationary, oscillating, and rotating cylinder. 
For a stationary case in an unsteady flow with a Reynolds number up to 200, a comparison of drag and lift coefficients was in perfect agreement with other results from the literature.
For moving objects, \cite{majumderReexaminingPartiallySaturatedcells2022a} demonstrated second-order accuracy for pressure, although the spatial order of accuracy in velocity shows degradation to first-order.
\cite{haussmannGalileanInvarianceStudy2020} studied vortex-induced vibrations for an eccentrically positioned cylinder in a transient Couette flow and reported the suitability of the PSM to simulate vortex-induced vibrations properly.
In \cite{trunkSimulationArbitrarilyShaped2018}, a stationary cylinder's drag and lift coefficients are presented and show good agreement with the literature.

\begin{figure}
	\centering
    \includegraphics[width=0.5\linewidth, trim={3cm 0cm 3cm 0cm},clip]{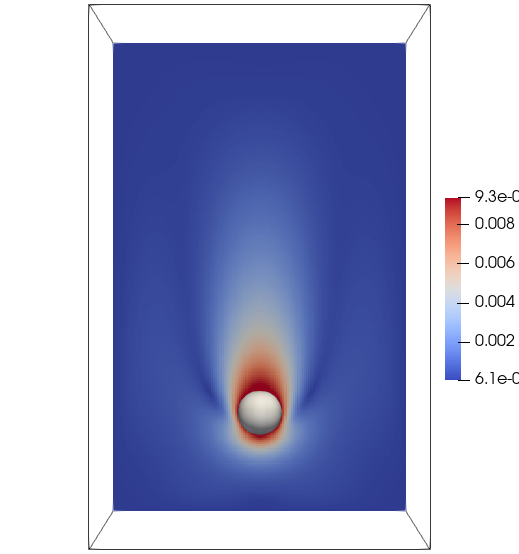}
	\caption{Visualization of the settling sphere validation test. The Sphere is positioned at the top of the domain and settles down due to gravitational force.}
	\label{fig:settlingSphere}
\end{figure}

\begin{figure}
	\centering
    \includegraphics[width=\linewidth, trim={0cm 0cm 1cm 0cm},clip]{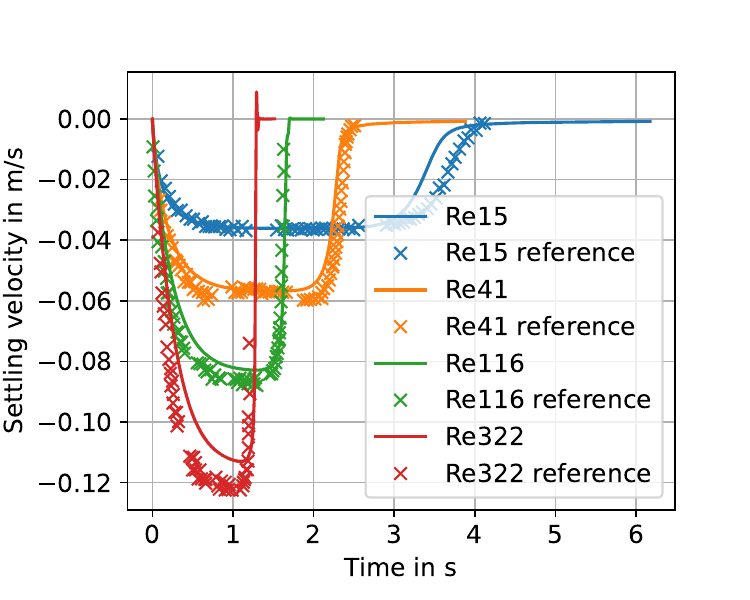}
	\caption{Settling velocities for the settling sphere validation case for different Reynolds numbers. Reference velocities from \cite {tencateParticleImagingVelocimetry2002}.}
	\label{fig:settlingSphereVel}
\end{figure}

To validate the accuracy of our fluid-structure interaction implementation of the PSM, we reproduce the experiments in \cite{tencateParticleImagingVelocimetry2002}. 
The validation case involves a sphere freely settling under the influence of gravity in different oils, as shown in \autoref{fig:settlingSphere}. 
This test case has become a standard for the validation of moving particle simulations (\cite{seilOnsetSedimentTransport2018}, \cite{biegertCollisionModelGrainresolving2017}, \cite{rettingerEfficientFourwayCoupled2022}), and is therefore suitable for testing the fluid-structure interaction of our approach.
For the oils that fill the domain of the test case, we study four different viscosities, resulting in the Reynolds numbers $Re=15$, $Re=41$, $Re=116$, and $Re=322$.
We set the domain size to $135 \times 135 \times 216$ LBM cells, with no-slip conditions on the boundaries in each direction. 
The PSM handles the sphere motion, and the mapping approach presented in \autoref{sec:mapping} is used.
We configured the PSM setup with an SRT collision operator and the solid collision operator SC2 from \autoref{equ:solid_collision2}.
In \autoref{fig:settlingSphereVel}, the settling velocity of the sphere over time is shown for the four different Reynolds numbers.
The maximum settling velocities for $Re=15$ and $Re=41$ agree perfectly with the reference from \cite{tencateParticleImagingVelocimetry2002}.
For the validation case with $Re=116$ and $Re=322$, the results are still in good agreement with the reference data with a relative error of 4\% and 7\%, respectively.
The velocity decrease upon hitting the bottom of the domain differs slightly from the reference points.
This can be explained since, here, no lubrication correction represents the unresolved interaction forces between the sphere and the wall when the distance is small.
However, this is not relevant in our case since we are only interested in modeling moving domains where the solid objects are not colliding.
Therefore, there is no need for a lubrication correction model.

\section{Performance Results}
\label{sec:performance}
\subsection{Node-Level Performance}
\label{sec:nodePerformance}

\begin{figure}
\begin{subfigure}{.2\textwidth}
  \centering
  \includegraphics[width=\linewidth, trim={14cm 1cm 14cm 1cm},clip]{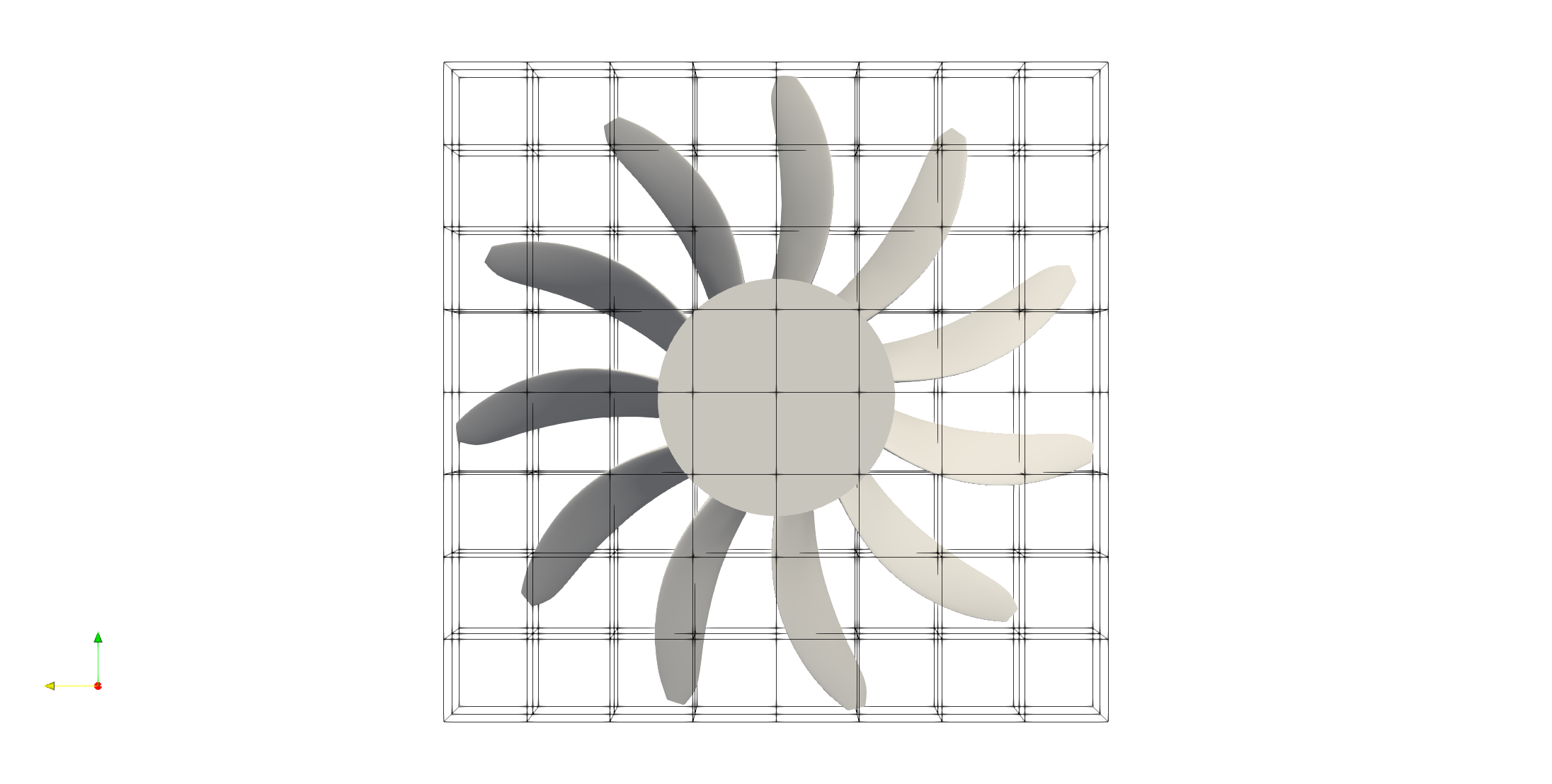}
  \caption{Scenario A: Geometry rotating over full domain}
  \label{fig:rotor_CPU}
\end{subfigure}%
\begin{subfigure}{.3\textwidth}
  \centering
  \includegraphics[width=\linewidth, trim={0cm 0cm 0cm 0cm},clip]{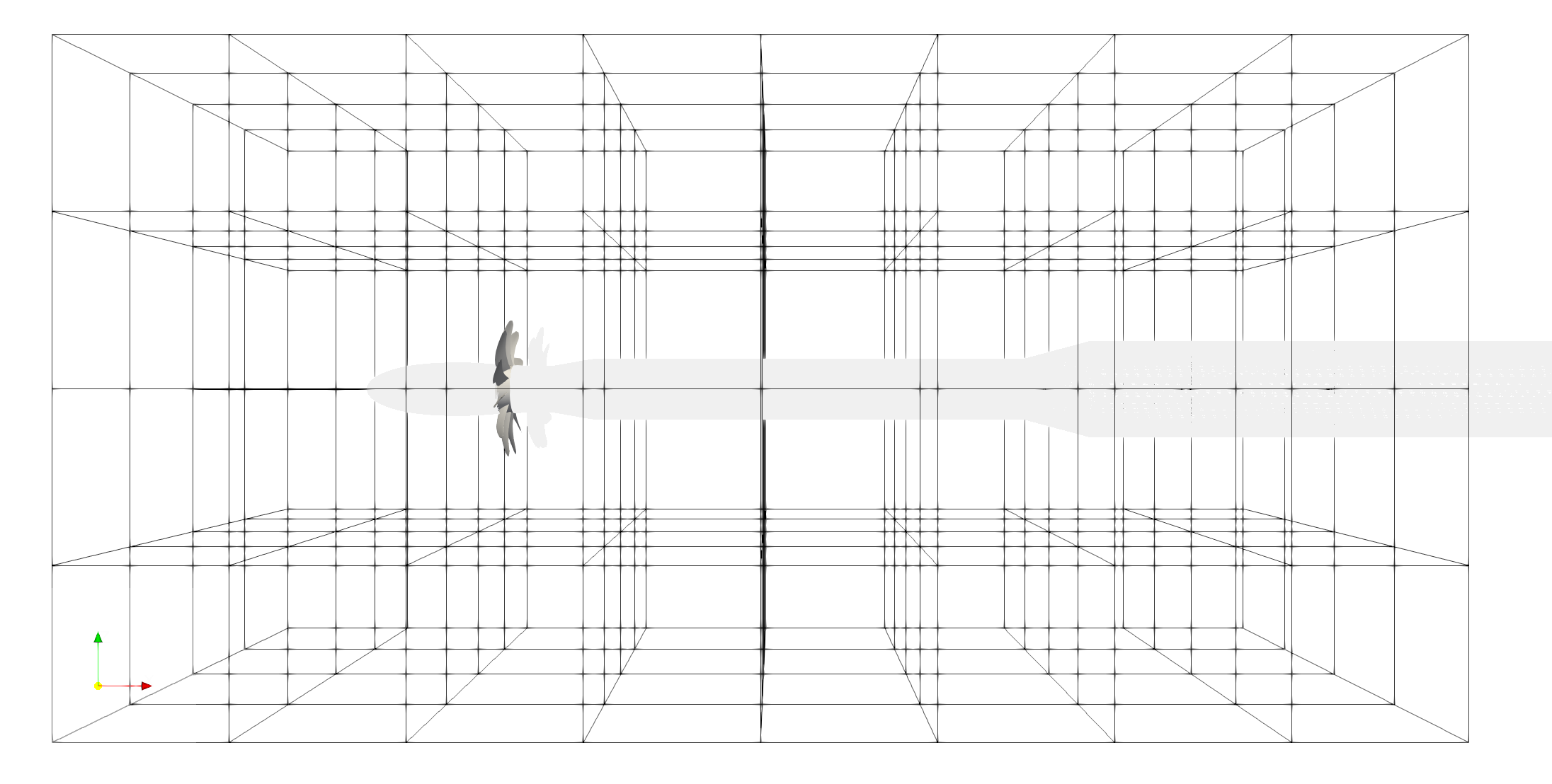}
  \caption{Scenario B: Geometry rotating in small part of domain}
  \label{fig:CROR_CPU}
\end{subfigure}
\caption{Block structure of test scenarios with 128 blocks ($64^3$ cells per block) for a CPU node-level performance benchmark on LUMI-C.}
  \label{fig:blockstructures_CPU}
\end{figure}

Besides the physical validation, we are also interested in the performance of the generated code from \autoref{sec:psm_theory} and the cost of the interpolation kernel presented in \autoref{sec:mapping}.
We start with a CPU node-level benchmark on the LUMI-C supercomputer on a single CPU node. 
The primary CPU node of LUMI-C consists of two AMD EPYC 7763 CPUs (2.45 GHz base, 3.5 GHz boost) with 64 cores each, so 128 cores per node. 
On this architecture, we tested the two scenarios shown in \autoref{fig:blockstructures_CPU}.
The first scenario is a rotating geometry over the whole domain. 
In contrast, the second scenario holds a rotating geometry in a small part of the domain, similar to the CROR test case setup in \autoref{fig:CROR_big_run}. 

We run the simulations using the massively parallel multi-physics framework \walberla{}, which provides an octree block structure. 
\walberla{} operates MPI parallel, and the simulation domain is decomposed into uniform blocks so that each MPI process operates on at least one block. 
These blocks contain the actual information for an LBM simulation, therefore storing and operating on one part of the LBM lattice.
The code generated by \lbmpy{} is executed on each lattice cell.
This workflow allows the generated efficient code to be executed on thousands of MPI processes.
For more information about the structure and functionality of \walberla{}, see \cite{waLBerla}, \cite{godenschwagerFrameworkHybridParallel2013}, \cite{schornbaumExtremeScaleBlockStructuredAdaptive2018} and the open-source code repository (\url{www.walberla.net}).

For the simulation domains in \autoref{fig:blockstructures_CPU}, both scenarios are divided into 128 blocks of $64^3$ LBM cells to fully utilize all available 128 CPU cores so that one core holds one block.
The LBM is configured with a D3Q19 stencil, a Cumulant collision operator, and the AA streaming pattern for all performance results below. 
For the PSM, we used the solid collision operator SC1 in \autoref{equ:solid_collision1} and restricted our fluid-structure interaction to a one-way coupling. 
This means that the geometry affects the flow behavior through \autoref{equ:PSM_rule}, but the fluid flow does not affect the geometry because we operate the rotors at a fixed rotation speed.

\begin{figure}
	\centering
     \includegraphics[width=\linewidth]{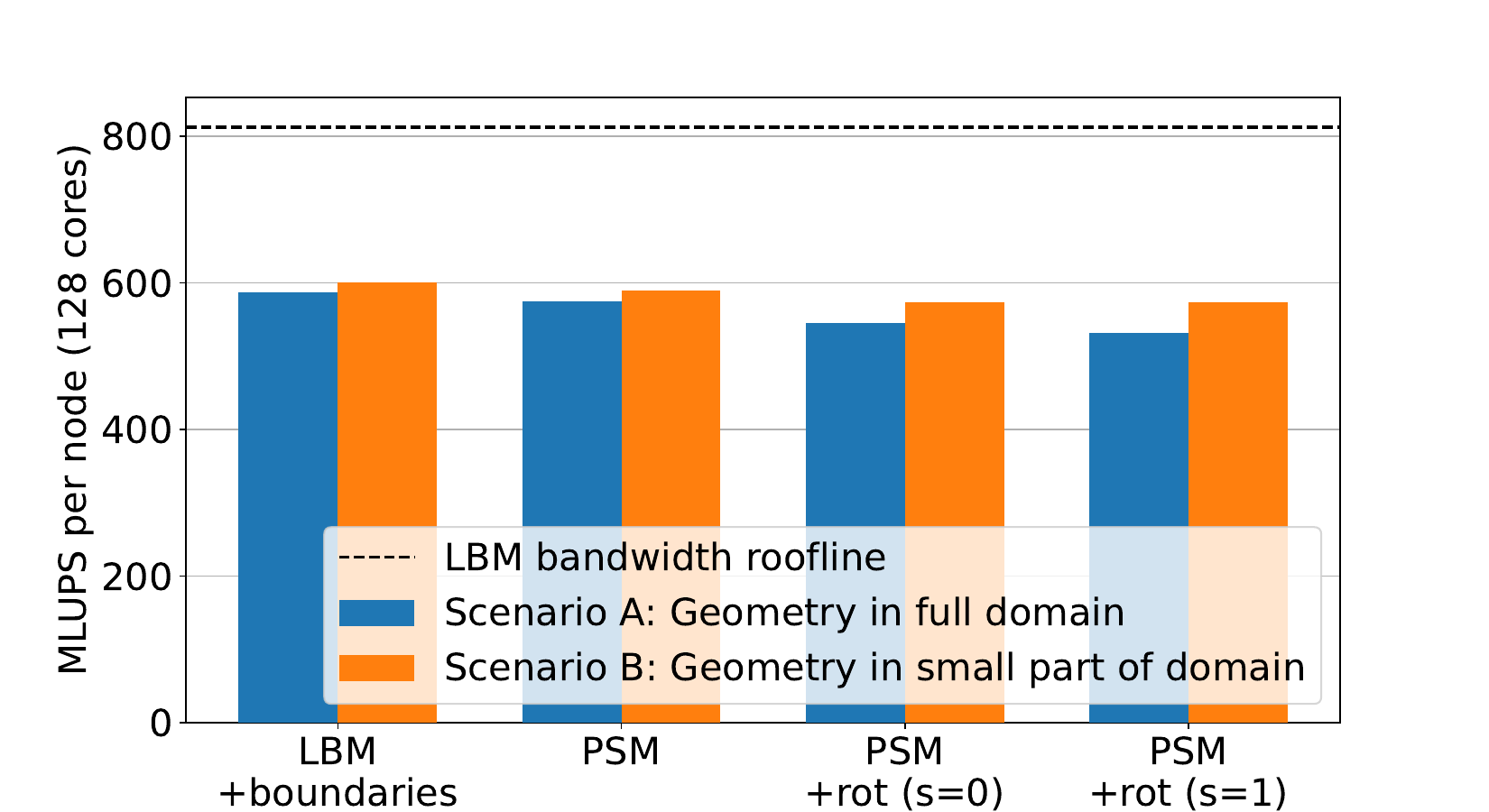}
    \caption{Node-level performance of PSM vs LBM on 2x AMD EPYC 7763 CPUs on LUMI-C with 128 blocks and $64^3$ cells per block. Scenarios A and B are visualized in \autoref{fig:blockstructures_CPU}.}
    \label{fig:Node_level_CPU}
\end{figure}

In \autoref{fig:Node_level_CPU}, we see the performance results for an LBM kernel compared to different PSM kernels. 
The communication routines and boundaries, such as inflow or outflow boundaries, are excluded in these node-level results to measure only the raw LBM performance plus the boundary handling of the geometry to make a fair comparison between LBM and PSM. 
We present the performance of our generated code in Mega Lattice Updates per Second (MLUPS), so this metric describes how many cell updates per second can be computed by the underlying architecture.
The LBM bandwidth roofline in \autoref{fig:Node_level_CPU} represents the theoretical peak performance of the LBM algorithm. 
We use the bandwidth roofline as a peak performance model because LBM algorithms are mostly memory-bound. 
A code is memory-bound when the arithmetical intensity of the code is low, such that the execution pipeline of the underlying architecture is stalled, waiting for data loads.
The LBM bandwidth roofline is calculated by dividing the CPU bandwidth by the number of memory accesses per LBM cell update.
With a stream benchmark, we measured a summed memory bandwidth of $\sim 247$ GB/s for the two CPUs on one LUMI-C node, and we used a D3Q19 stencil for the benchmarks. 
To update an LBM cell, we need two memory accesses per stencil direction, so one read of the PDFs and one write to store the PDFs, when using an in-place streaming pattern (\cite{wittmannComparisonDifferentPropagation2013}).
So for the bandwidth roofline we get $P_{\text{max}} = 247.000 \text{ MB/s} / (2\cdot 19\cdot 8 \text{ B}) = 812.5 \text{ MLUPS}$.

The first result on the left in \autoref{fig:Node_level_CPU} is a simple LBM kernel plus the no-slip boundary treatment for the rotor geometry. 
In this benchmark result, no rotation / translation of the geometry is handled, yet it remains stationary in the flow. 
First of all, we observed high bandwidth utilization. 
With 600.45 MLUPs for scenario A and 587.38 MLUPS for scenario B, we reach over 72\% of the theoretical peak performance for both scenarios. The difference between scenarios A and B is also very small. 
This is because the no-slip boundary handling for scenarios A and B takes only 1.78\% and 0.05\% of the run time, respectively.
The slight difference is due to the higher number of boundary cells in the simulation domain of scenario A. 

Next, we look at the PSM kernel from \autoref{sec:psm_theory}, which handles fluid and boundary cells in a single kernel. 
The geometry is still stationary.
Compared to the LBM kernel, we only lose about 2\% of the MLUPS when using the PSM, which is due to the additional memory access for reading the fraction field and the solid object velocity field in \autoref{equ:PSM_rule}.

Finally, we look at the performance of the PSM with an actual rotating geometry. 
The rotor in scenarios A and B rotates clockwise around its center.
The rotation is handled by building a new solid volume fraction field $B(\textbf{x},t)$ from the geometry field at each time step as explained in \autoref{sec:mapping}. 
Since the geometry field is initially populated in a preprocessing step, only the interpolation of the geometry field to the solid volume fraction field $B(\textbf{x},t)$ is performance-relevant. 
Here, we tested two super-sampling factors, $s=0$ and $s=1$, which should lead to a different workload.
In \autoref{fig:Node_level_CPU}, we show that the handling of the rotation (interpolation) kernel has only a tiny impact on the overall performance of the simulation. 
For scenario A, which is the worst case in terms of performance, because the rotating rotor covers a large part of the domain, the rotation handling reduces the performance by only 7\% 
The result is even better for scenario B and $s=1$, where the rotation handling takes up less than 3\% of the runtime. Furthermore, the super-sampling factor has a negligible impact on the performance of the rotation handling.
For an industrial application such as the CROR in \autoref{fig:CROR_geo}, the domain setup is close to scenario B, so the performance loss due to the rotation handling is expected to be less than 3\% of the run time.

\begin{figure}
	\centering
     \includegraphics[width=\linewidth]{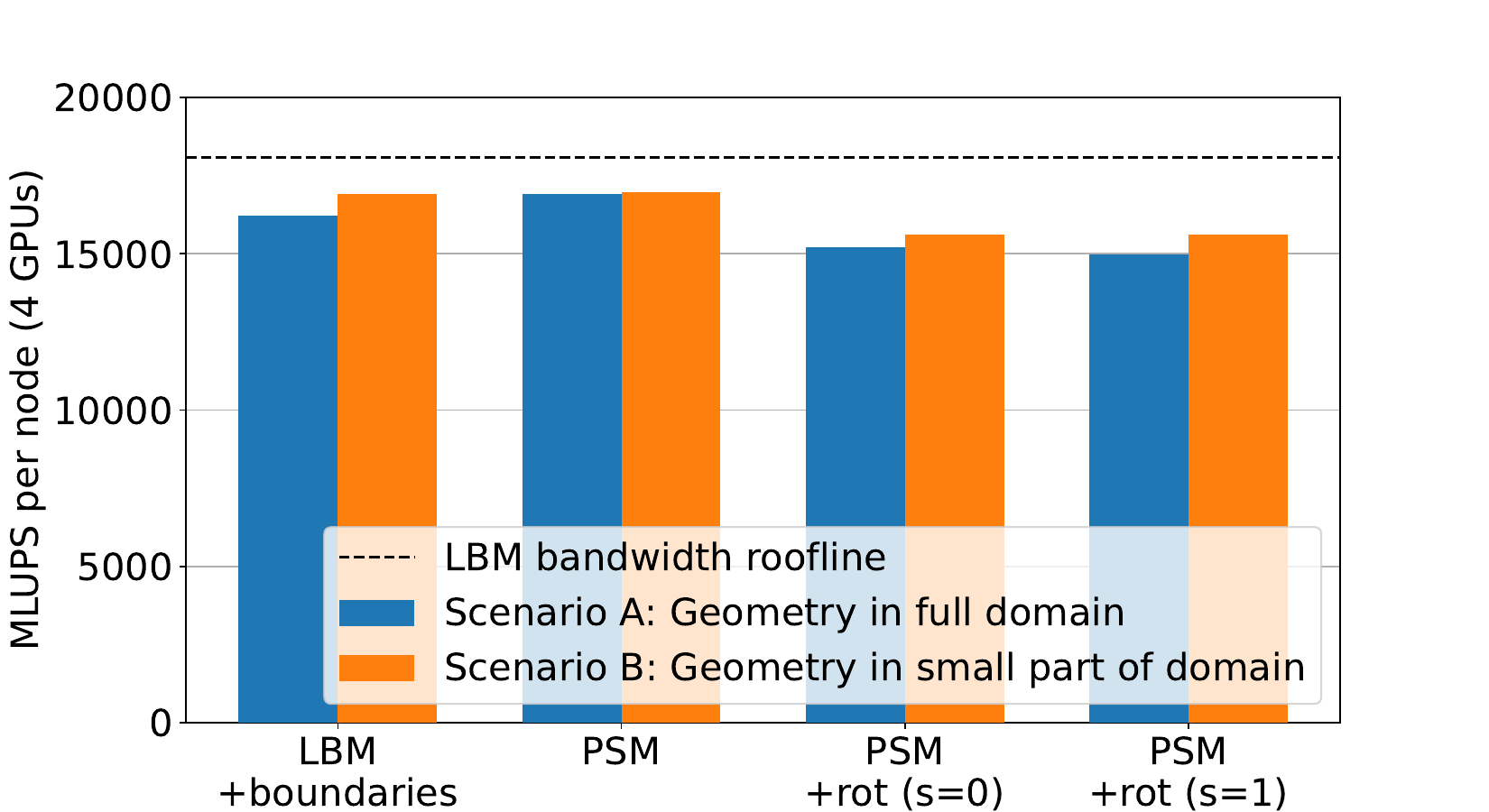}
    \caption{Node-level performance of PSM vs LBM on 4 NVIDIA A100 GPUs on JUWELS-Booster with 1 block per GPU and $256^3$ cells per block. Scenarios A and B are visualized in \autoref{fig:blockstructures_CPU}, but with a domain decomposition of only four blocks.}
    \label{fig:Node_level_GPU}
\end{figure}

For the GPU node-level performance, we studied the same two scenarios from \autoref{fig:blockstructures_CPU} and the same four kernel variants as for CPU node-level performance, but this time on a JUWELS booster node consisting of four \nvidia{} A100 40GB GPUs.
We have decomposed the domain into four blocks with $256^3$ cells per block so that one GPU handles one block and the GPU is fully utilized. 
The node-level performance results are presented in \autoref{fig:Node_level_GPU}.
Again, we introduce the LBM bandwidth roofline to estimate the theoretical maximum performance for our LBM kernel. It is calculated from the measured bandwidth of $4 \cdot 1366 \text{ GB/s}$ for the four GPUs, divided by the number of memory accesses to get a maximum performance of 18092 MLUPS.
The LBM kernel for scenario B with a small number of boundary cells achieves 16916 MLUPS, which is over 93\% of the maximum achievable performance. 
Scenario A is 5\% slower due to the domain's higher number of boundary cells.
We observe the same performance for the PSM kernel without rotation as for the LBM kernel for scenario B and even a speedup for scenario A. 
The higher number of boundary cells does not affect the superior performance of the PSM code.
Handling the geometry rotation shows a performance impact slightly higher than on the CPU, with a 10\% performance loss for scenario A and an 8\% performance loss for scenario B. 
Again, a higher super-sampling factor has no additional cost in terms of MLUPS. 

We conclude that we have managed to implement a handling for complex moving geometries on CPUs and GPUs that introduces only a low performance penalty when using PSM compared to a pure LBM algorithm, with the difference that the pure LBM step can not handle the rotation/translation of the geometry. 
Extending the simple LBM to moving boundaries would require, among others, complex PDF reconstruction algorithms, which are completely avoided using the PSM.

\subsection{Scaling Efficiency}

We have shown that the single-node performance of PSM plus rotation overhead is excellent. 
Next, the scaling efficiency of our code is investigated. 
To do so, we set up a simulation for the CROR industrial case that runs on the LUMI-C CPU and the JUWELS Booster GPU clusters. 
These measurements perform a complete simulation run, which includes communication and boundary routines in addition to the PSM and rotation steps.
We also included mesh refinement for the CPU runs. 
Mesh refinement is used to resolve flows in regions of interest efficiently, so for the example of the CROR case, we refine the area around the rotor blades.
With this approach, we avoid increasing the resolution of the whole lattice, but only in specific regions, which can be very beneficial regarding computational costs. 
The mesh refinement approach in \walberla{} provides good accuracy even for highly turbulent flows, as shown in \cite{holzerCodeGenerationLattice2025}.

\begin{figure}
	\centering
     \includegraphics[width=\linewidth]{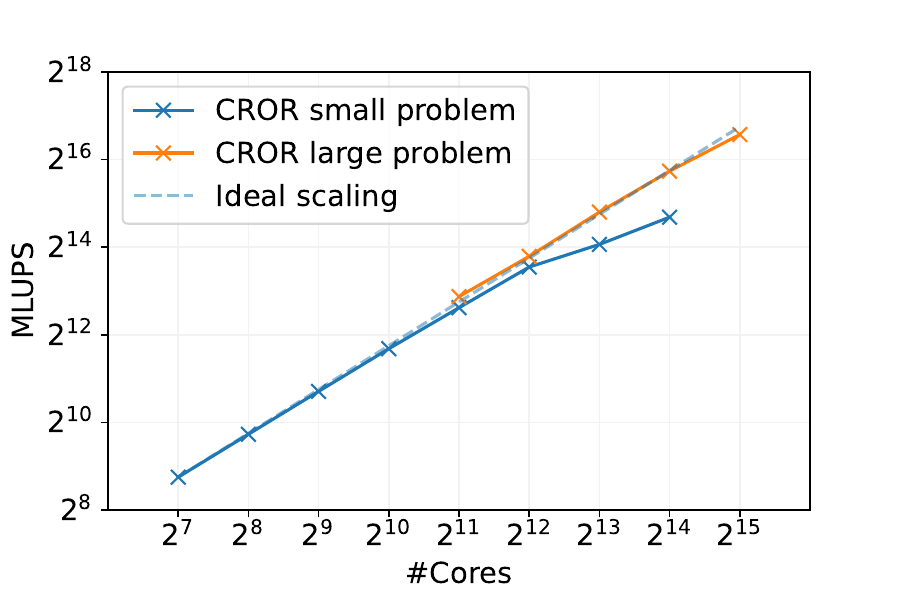}
    \caption{Strong scaling of the CROR simulation on the CPU partition of LUMI  (AMD EPYC 7763 CPU) for the two problem sizes in \autoref{tab:scaling}.}
    \label{fig:scaling_CPU}
\end{figure}

We present the parameters for four strong scaling runs in \autoref{tab:scaling}.
On the CPU partition of LUMI, we measured a medium-sized run with $2.15 \cdot 10^8$ LBM cells and four levels of refinement to end up with a refined cell size of 0.001 m in the region of interest. 
We start the strong scaling on a single node of LUMI-C, so on 128 cores, and take the resulting performance of 430.582 MLUPS as the baseline for the linear scaling curve in \autoref{fig:scaling_CPU}. 
The strong scaling performance on a single node is about 30\% lower than the single-node measurements in \autoref{fig:Node_level_CPU} because here we also include the boundary handling for inflow and outflow as well as communication routines and the mesh refinement approach in our measurements. 
In \autoref{fig:scaling_CPU}, we observe that we achieve an excellent scaling efficiency of 86\% on up to $2^{12} = 4096$ cores (32 nodes). 
However, above 32 nodes, the scaling efficiency drops sharply. 
This is because the problem size no longer matches the number of processes.
For 64 nodes, each CPU core operates on less than two blocks, and for 128 nodes, it operates on less than one block. 
Since the block size is small, with only $24^3$ cells per block, the cores are not fully utilized and spend significant time waiting for data communication from neighboring processes. 
Also, as noted by \cite{schornbaumBlockStructuredAdaptiveMesha}, the workload per core may not be perfectly balanced with a small number of blocks per core.

To increase the CPU usage on a large number of nodes, we set up a second strong scaling on LUMI. 
This time, the domain of the CROR simulation is resolved with $3.53 \cdot 10^9$ LBM cells and five refinement levels to end up with the finest grid resolution of 0.0003 m at the rotor blades.
The strong scaling starts on 16 nodes, as the required memory for the larger case exceeded the memory available at lower node counts.
For the larger problem size, we present in \autoref{fig:scaling_CPU} a scaling efficiency of 81\% on up to 256 CPU nodes, corresponding to 32768 CPU cores, when using the first strong scaling result of the large problem on 16 nodes as a baseline for the linear scaling.
Using the small-scale run on one node as a baseline, a scaling efficiency of 88\% on 32768 CPU cores is achieved.

\begin{table}
\caption{Domain setups for the strong scaling experiments on LUMI-C and JUWELS Booster.}
\label{tab:scaling}
\vspace{-0.3cm}
\begin{tabular}{ |c||c|c||c|c| } 
\hline
 &  \multicolumn{2}{c||}{CPU (LUMI-C)} & \multicolumn{2}{c|}{GPU (JUWELS Booster)} \\ \hline \hline  
Problem size (\#cells) &  2.15E08 & 3.53E09 & 2.68E08 & 4.29E09 \\ \hline
Blocks count & 15598 & 255710 & 64  & 1024 \\ \hline
Cells per block & $24^3$ & $24^3$  & $256\cdot128^2$ & $256\cdot128^2$ \\ \hline
Refinement levels & 4 & 5 & 0 & 0 \\ \hline
Refined cell size (m) & 0.001 & 0.0003 &  0.004 &  0.0023 \\ \hline
\end{tabular}
\end{table}

The scaling efficiency of the CROR on the GPU partition of the JUWELS supercomputer was also investigated. 
Similar to the CPU scaling experiments, we set up two problem sizes to span the scaling range from one node to the maximum number of assignable nodes on the cluster, which is 256 nodes and, thus, 1024 GPUs on the JUWELS booster (see \autoref{tab:scaling}). 
The small case with  $2.68 \cdot 10^8$ LBM cells is comparable to the small case on CPU. 
For the GPU strong scaling, we excluded mesh refinement because the proper use of mesh refinement requires small block sizes. 
Small block sizes inherently reduce the performance on GPUs because they lead to small kernel calls and under-utilization of the hardware (\cite{holzerCodeGenerationLattice2025}). 
Therefore, we use a block size of $256 \cdot 128^2$ LBM cells per block for the GPU measurements.
Work is in progress for the \walberla{} framework to increase the performance of the mesh refinement on GPU architectures.

\begin{figure}
	\centering
     \includegraphics[width=\linewidth]{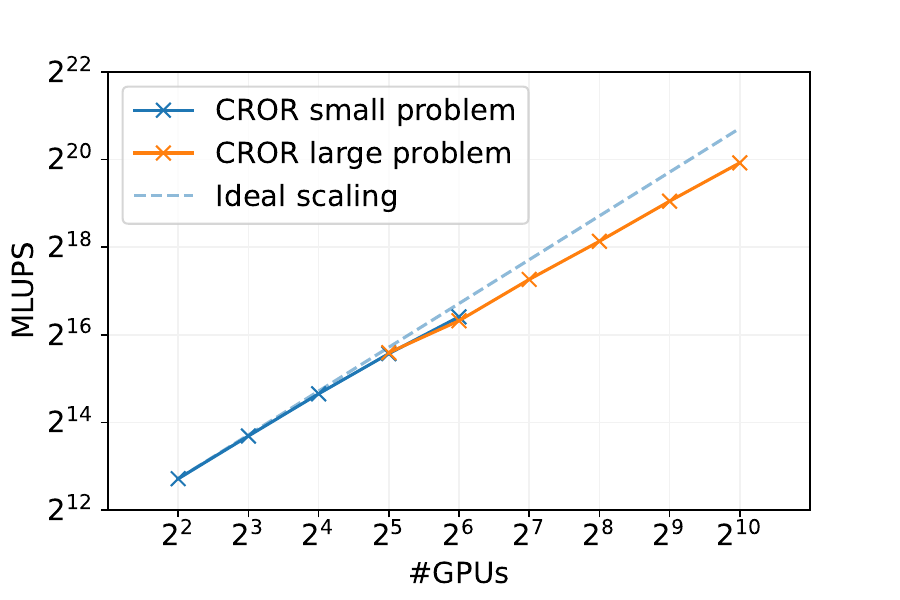}
    \caption{Strong scaling of the CROR simulation on the GPU partition of JUWELS (NVIDIA A100 GPUs) for the two problem sizes in \autoref{tab:scaling}.}
    \label{fig:scaling_GPU}
\end{figure}

In \autoref{fig:scaling_GPU}, we present the strong scaling results for the CROR simulation on GPU. 
We achieve a scaling efficiency of 81\% on up to 64 GPUs (16 nodes). 
We cannot scale this problem size further because, for more than 64 GPUs, there would be less than one block per GPU. 
On the other hand, the number of blocks cannot be increased because, therefore, the number of cells per block has to be reduced. 
Again, this leads to underutilization of the GPUs, thus resulting in insufficient performance results.

Therefore, a larger problem size is set up to scale up to 1024 GPUs. 
The large problem size contains $4.29 \cdot 10^9$ LBM cells and is limited to running on at least 32 GPUs, not to exceed the memory. 
We perform strong scaling for up to 1024 GPUs on the JUWELS Booster system and achieve up to 996 Giga Lattice Updates, or nearly one Tera Lattice Update per second. 
This means that the generated PSM code can update all $4.29 \cdot 10^9$ LBM cells 232 times per second, so it needs only 0.0043 seconds per time step for the CROR simulation.
The scaling efficiency on 1024 NVIDIA A100 GPUs is 63\% compared to the large problem's performance on 32 nodes. 
The excellent scaling properties of \walberla{} are already demonstrated in other publications (\cite{schornbaumBlockStructuredAdaptiveMesha}, \cite{holzerCodeGenerationLattice2025}).
To achieve better scalability for these GPU measurements, we need to increase the block size further to achieve better utilization of the GPUs and to make the communication hiding strategies in \walberla{} more efficient.

\begin{figure}
\end{figure}

\section{Conclusion and Outlook}
\label{sec:outlook}
This work presents the steps towards large-scale and physically accurate fluid flow simulations around complex moving geometries such as the CROR. 
The present study has implemented the PSM and an efficient approach for mapping the complex geometry onto the LBM lattice. 
The PSM and the mapping approach were validated using a settling sphere test case, resulting in good agreement with reference data. 
The node-level performance of the PSM approach, including the handling of the rotation of the geometry on CPU and GPU architectures, was studied, and excellent performance results were obtained. 
A simulation was configured for the CROR geometry, employing a mesh resolution of 0.3 millimeters around the area of the rotor blades. 
The application demonstrated efficient scaling on up to 32,768 CPU cores and 1,024 NVIDIA A100 GPUs. 

Additionally, a visualization was executed on 36 NVIDIA A100 GPUs using the JUWELS Booster.
The result is displayed in \autoref{fig:CROR_big_run}. 
The Q-criterion of the flow field around the CROR is presented for a Reynolds number of 4640 after 0.21 seconds of real-time, with a rotor blade rotation of 73.8 rad/s.
The implementation of the PSM approach has yielded an efficient and stable simulation. 

In order to achieve the final objective of a relevant simulation of the CROR, additional steps must be taken.
A proper validation of the physical accuracy of the PSM and the proposed mapping approach for complex geometries should be performed. 
However, it is challenging to locate experimental data in this area.
Also missing is a validation for the PSM in highly turbulent flow, such as a test case of a turbulent flow around a sphere from \cite{holzerCodeGenerationLattice2025}.

To get a physically meaningful simulation of the CROR in air, it is necessary to drastically increase the Reynolds number to the order of $10^6$. 
Consequently, it is crucial to enhance the resolution close to the geometry by extensively employing mesh refinement techniques. 
Furthermore, wall models and turbulent inflow conditions could be considered to achieve an appropriate simulation of the CROR in turbulent flow.
Finally, a comparison of the simulation results with experimental data from the CROR in wind tunnels is desirable.

\section*{Acknowledgements}
This work was supported by the SCALABLE project (\url{www.scalable-hpc.eu}). This project has received funding from the European High-Performance Computing Joint Undertaking (JU) under grant agreement No 956000. The JU receives support from the European Union’s Horizon 2020 research and innovation program and France, Germany, and the Czech Republic. 
This project has also received funding from the European High Performance Computing Joint Undertaking under grant agreement n°101144014.
The authors gratefully acknowledge the Gauss Centre for Supercomputing e.V. (\url{www.gauss-centre.eu}) for funding this project by providing computing time through the John von Neumann Institute for Computing (NIC) on the GCS Supercomputer JUWELS at Jülich Supercomputing Centre (JSC). 
We acknowledge the EuroHPC Joint Undertaking for awarding this project access to the EuroHPC supercomputer LUMI, hosted by CSC (Finland) and the LUMI consortium through a EuroHPC Regular Access call. 
The authors gratefully acknowledge the scientific support and HPC resources provided by the Erlangen National High Performance Computing Center (NHR@FAU) of the Friedrich-AlexanderUniversität Erlangen-Nürnberg (FAU).

\section*{Author contributions}
\noindent \textbf{P. Suffa:} Conceptualization, Methodology, Software, Validation, Formal analysis, Investigation, Data Curation, Writing - Original Draft, Visualization, Project administration \\
\textbf{S. Kemmler:} Conceptualization, Software, Writing - Review \& Editing \\
\textbf{H. Koestler:} Resources, Writing - Review \& Editing, Supervision, Funding acquisition \\
\textbf{U. Ruede:} Writing - Review \& Editing, Supervision, Funding acquisition \\

\section*{Conflict of Interest Statement}
The authors have no conflicts to disclose.

\section*{Data Availability Statement}
The data that support the findings of this study are openly available on Zenodo: \url{https://doi.org/10.5281/zenodo.14801403}

\nocite{*}
\section*{References}

\providecommand{\noopsort}[1]{}\providecommand{\singleletter}[1]{#1}%

\end{document}